# Crystal orientation and detector distance effects on resolving pseudosymmetry by electron backscatter diffraction

Edward L. Pang, Christopher A. Schuh

*Department of Materials Science and Engineering, Massachusetts Institute of Technology, 77 Massachusetts Ave, Cambridge, MA 02139, USA*

*Corresponding author. Email address: schuh@mit.edu (C.A. Schuh)

**Abstract**

Accurately indexing pseudosymmetric materials has long proven challenging for electron backscatter diffraction. The recent emergence of intensity-based indexing approaches promises an enhanced ability to resolve pseudosymmetry compared to traditional Hough-based indexing approaches. However, little work has been done to understand the effects of sample position and orientation on the ability to resolve pseudosymmetry, especially for intensity-based indexing approaches. Thus, in this work we quantitatively investigate the effects of crystal orientation and detector distance in a model tetragonal $ZrO_2$ (c/a=1.0185) material. We identify orientations that are easiest and most difficult to correctly index, characterize the effect of detector distance on indexing confidence, and analyze these trends based on the appearance of specific zone axes in the diffraction patterns. Our findings also point to the clear benefit of shorter detector distances for resolving pseudosymmetry using intensity-based indexing approaches.

# 1. Introduction

Electron backscatter diffraction (EBSD) has become a widespread tool to rapidly characterize crystal orientations and microstructures of crystalline materials (Schwartz *et al.*, 2009). Traditionally, EBSD patterns are indexed by identifying bands using the Hough transform and then comparing inter-band angles to a lookup table based on the crystal structure to deduce the orientation. In pseudosymmetric materials, for which multiple crystallographically distinct orientations give rise to similar diffraction patterns because of a near symmetry element (Nowell & Wright, 2005), it is difficult to obtain unambiguous indexings using Hough-based indexing schemes; often, inverse pole figure (IPF) maps exhibit a "checkerboard" pattern, showing that the algorithm cannot decide on a single orientation (Ryde, 2006; Wright, 2006; Ocelík *et al.*, 2017). The past few years have seen the development of a new family of intensity-based indexing approaches that perform full image comparisons with high-quality dynamically simulated patterns (Chen *et al.*, 2015; Nolze *et al.*, 2017; Foden *et al.*, 2019). These intensity-based indexing approaches are more sensitive to subtle pattern differences and can thus more easily distinguish between pseudosymmetric crystal orientations compared to Hough-based indexing schemes (Nolze *et al.*, 2016; Jackson *et al.*, 2018; De Graef *et al.*, 2019; Lenthe *et al.*, 2019; Pang *et al.*, 2020*b*).

Yet, much is still unknown about what affects the ability of intensity-based indexing methods to distinguish between pseudosymmetry variants. Recently, Lenthe et al. developed an intensity-based autocorrelation method over the entire Kikuchi sphere to identify potential pseudosymmetry rotation elements (Lenthe *et al.*, 2019). Here, we instead ask the complementary question of how our ability to distinguish between already known pseudosymmetry rotations is affected by which portions of the Kikuchi sphere are captured in a measured pattern. Previously, Nowell and Wright showed that the ability of Hough-based indexing to resolve pseudosymmetry issues was dependent on crystal orientation in a number of different materials (Nowell & Wright, 2005). It makes sense that intensity-based indexing approaches would also be sensitive to crystal orientation, but this has yet to be studied in detail in any material. Recently, we tested our modified dictionary indexing method on a tetragonal zirconia ($ZrO_2$) material and found a distribution of confidence index (CI) values, here defined as the difference in normalized dot product (NDP) between the variants with the highest and second-highest

NDP, over the 100 random orientations tested (Pang *et al.*, 2020*b*). This observation suggested that pseudosymmetry was more easily resolved for certain crystal orientations, but this issue was not systematically investigated in that work.

Another parameter of interest is the detector distance from the sample; for Hough-based indexing, it is generally believed that a shorter detector distance aids in resolving pseudosymmetry because a larger solid angle is captured by the detector (Ryde, 2006; Nowell & Wright, 2005). However, there has been conflicting experimental evidence regarding this point. Whereas Nowell and Wright found in alumina that indexing confidence consistently decreased at shorter detector distances (Nowell & Wright, 2005), Karthikeyan et al. found the opposite in a ferritic steel and also suggested that shorter detector distances came with a drawback of reduced orientation accuracy (Karthikeyan *et al.*, 2013). Both of these points require additional confirmation, and these effects still need to be investigated and quantified for intensity-based indexing approaches.

Thus, the position and orientation of the sample significantly impact the ability to differentiate pseudosymmetric variants, and in this work we systematically study those effects in the specific context of intensity-based indexing approaches in a tetragonal $ZrO_2$-13.5$CeO_2$ (c/a=1.0185) material. In Section 2, we characterize the orientation dependence of CI and reveal crystal orientations that are easiest (or most difficult) to distinguish from their pseudosymmetric counterparts. In Section 3, we subsequently quantify the effect of detector distance on CI and orientation accuracy using a modified dictionary indexing method. In both sections, we reveal the underlying causes of both effects by analyzing characteristic zone axes in the EBSD patterns.

**2. Orientation dependence of the ability to resolve pseudosymmetry**

*2.1. Sample vs. detector normal*

To investigate if there is a systematic orientation dependence on our ability to resolve pseudosymmetry, here we more carefully analyze the results from the 100 simulated patterns of random orientations of tetragonal $ZrO_2$-13.5$CeO_2$ indexed using a modified dictionary indexing method in (Pang *et al.*, 2020*b*). These patterns of size 480×480 px were simulated with *EMsoft 4.0* using a 70° sample tilt, 5° camera elevation, detector pixel size of 50 μm, 25 kV accelerating voltage, and pattern center ($X^*$, $Y^*$, $Z^*$) of (0 px, 80 px, 15000 μm) in the *EMsoft* coordinate system (Jackson *et al.*, 2019). More detailed parameters are given in (Pang *et al.*, 2020*b*). These are the same parameters used for all simulated patterns in this paper unless specified otherwise.

We plot the CI values from a 1×1 binned noise-free dataset in an inverse pole figure (IPF) of the sample normal direction in Fig. 1a. The pattern with the highest CI of 0.084 had a surface normal near the [011] crystal direction. Most of the points with high CI are in this general region, but there are also a number of points with low CI interspersed throughout, which is confusing. The pattern with the lowest CI of 0.016 was found to have a surface normal near [041]. Again, nearby points generally appear to have CI values in the lower half of the range, but the orientation dependence is a bit unclear.

At this point, we note that the symmetry of the pattern at the detector depends on the zone axes that impinge upon it, not the zone axes that are normal to the sample surface. This suggests that rather than analyzing the IPF of the sample normal direction as is typically done, we should instead analyze the IPF of the detector normal direction, which would be independent of the relative positioning of the sample and detector. This new IPF is plotted in Fig. 1b, which reveals a clearer orientation dependence. Now, the points with high CI appear to be clustered around the [001] direction, and points with low CI are clustered towards the center of the IPF.

The pattern with the highest CI is shown in Fig. 1d, which reveals that the main zone axis in the center of the pattern is $[00\bar{1}]$, the c-axis. The indexing of this zone axis for the best pseudosymmetric solution is $[\bar{1}10]$, which corresponds to the a-axis of the pseudo-cubic double cell (Fig. 1c). The diffraction pattern reflects the symmetry of the crystal down the four-fold c-axis, whereas when it is exchanged with the a-axis, the symmetry breaks down into a two-fold axis. The intensity difference plot between the pattern and that of its pseudosymmetric variant, shown in Fig. 1e, reflects this, as there are clear differences in intensity that arise from the distortions corresponding to the symmetry breaking. This reveals why orientations where the [001] axis of the crystal is nearly parallel to the detector normal are most easily distinguished. Note that here and throughout this paper, we show circular patterns because a circular mask was used during indexing that only considers the displayed portion of the pattern.

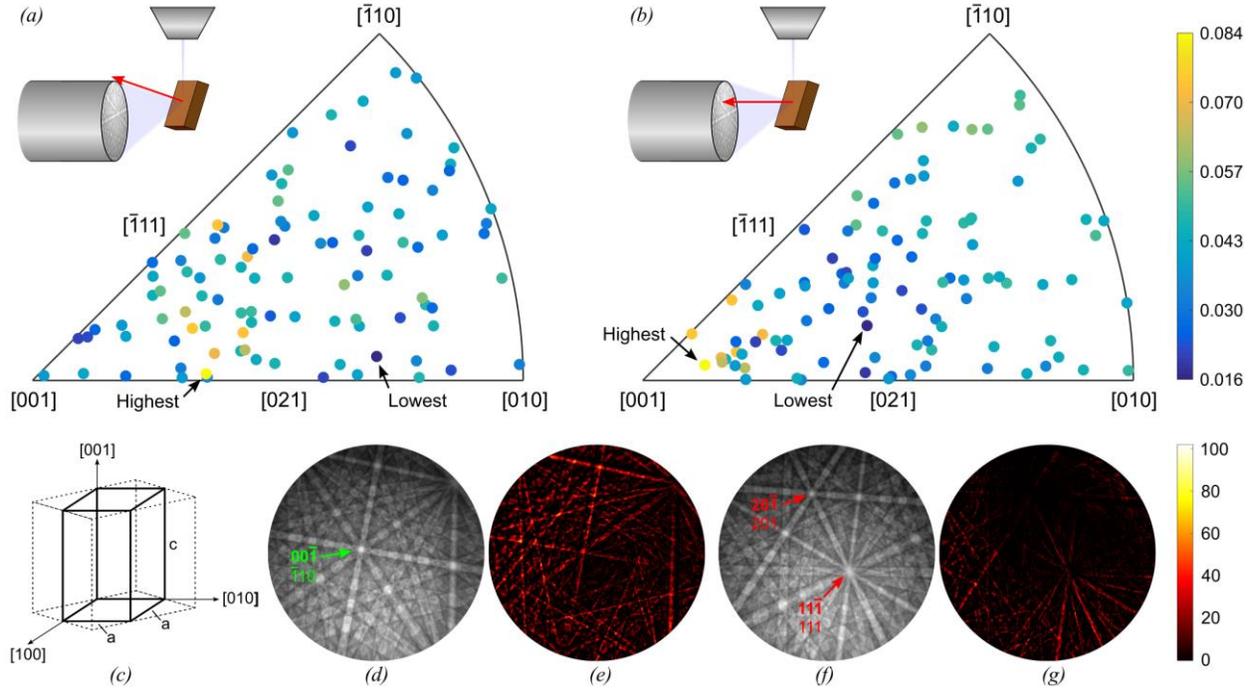

**Fig. 1.** Orientation dependence of pseudosymmetry in tetragonal $ZrO_2$-13.5$CeO_2$. (a-b) IPF map with points colored by CI, in reference to the (a) sample normal and (b) detector normal directions as shown in the schematics. (c) Illustration of the primitive unit cell (solid lines) in which all indexings are given and the non-primitive double cell in which pseudosymmetry is apparent (dashed lines). (d-e) Pattern with the highest observed CI and intensity difference with that of the best-matching pseudosymmetric orientation. (f-g) Pattern with the lowest observed CI and intensity difference with that of the best-matching pseudosymmetric orientation. For parts d and f, the top label in bold is the correct indexing of the zone axis, and the lower label is the indexing for the best-matching pseudosymmetry variant.

In contrast, the pattern with the lowest CI is shown in Fig. 1f. The two most prominent zone axes of the correctly indexed solution are $[20\bar{1}]$ and $[11\bar{1}]$, but these zones are indexed as the symmetry related $[201]$ and $[111]$ axes in the incorrect pseudosymmetric solution. This therefore explains why pseudosymmetry is so difficult to resolve in this pattern, as the primary features of the pattern exhibit the same symmetry, and it is only in the other less prominent parts of the pattern that distinguishing differences arise. This is confirmed by the difference plot shown in Fig. 1f, which reveal only small intensity differences near the edges of the pattern. This confirms why points in the region between the $[\bar{1}11]$ and $[021]$ directions in the detector normal IPF (Fig. 1b) have a low CI, as these patterns feature two prominent zone axes with identical symmetry between the top two solutions. The fact that these patterns can even be distinguished highlights the value of using the intensity information in the patterns via a pattern matching type approach as opposed to Hough-based indexing.

*2.2. Full orientation dependence*

It is often assumed that rotating an EBSD pattern, corresponding to a different crystal orientation attained by a rotation about the detector normal, has little effect on the ability to resolve pseudosymmetry. This is implicitly assumed when only a single IPF is used to represent the crystal orientation such as in Fig. 1b or in (Nowell & Wright, 2005). The fact that a trend can be seen in Fig. 1b does suggest that crystal rotations about the detector normal have a secondary effect compared to the detector normal direction in the crystal; however, some level of scatter is clearly present. We can evaluate this more clearly by creating a uniform grid of 981 crystal orientations in the detector normal IPF space and then rotating each point about the detector normal direction by angles of 0, 60, 120, 180, 240, and 300°. Each of these six rotations gives a unique crystal orientation, even though all six points share the same detector normal direction and thus appear at the same location in the detector normal IPF. We then simulate EBSD patterns for all 5886 orientations and calculate the CI value for each orientation from NDP values that are obtained using *EMsoft*'s *EMFitOrientation* program (we used a maximum step size of 0.03 in homochoric space, a circular mask with diameter equal to the image width, high-pass filter parameter of 0.05, and adaptive histogram equalization with 10 regions) to optimize the crystal orientation. This orientation optimization step is necessary because the orientation giving the maximum NDP usually slightly deviates from the ideal pseudosymmetric orientation, obtained for tetragonal $ZrO_2$ by a 90° rotation about [110] or [1$\bar{1}$0]; omitting this step would overestimate the CI.

Fig. 2 shows the detector normal IPF for the minimum, mean, and maximum CI values of the six rotations at each grid point. While there are clear differences in the quantitative CI values with varying rotations about the detector normal, the general trend is the same in all cases, with the highest CI values near [001] and the lowest CI values between [$\bar{1}$11] and [021], in agreement with Fig. 1. Curiously, the highest CI orientations appears to lie on a band slightly away from [001]. To understand why these orientations have the highest CI, in Fig. 3 we take a closer look at some of these patterns.

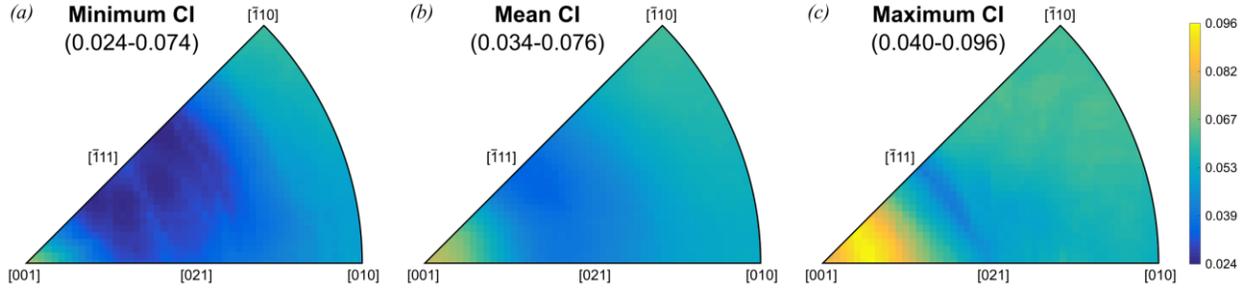

**Fig. 2.** Inverse pole figures in the detector normal direction showing the (a) minimum, (b) mean, and (c) maximum CI observed over the rotations tested. Pattern center is located at $X^*=0$ px, $Y^*=80$ px, and $Z^*=15000$ μm.

Fig. 3a shows the pattern for the [001] orientation, which has a CI of 0.0795. Towards the edges of the pattern, two crystallographically ambiguous <111> zones are visible. However, more prominently displayed in this pattern is the crystallographically unique [001] zone that lies slightly above the middle of the pattern; in fact, it is located at the "pattern center" (which we will term the "projection center" throughout this work to avoid confusion when describing locations of zone axes within patterns) of $X^*=0$ px and $Y^*=80$ px (or $X^*=0.5$ and $Y^*=0.666667$ in fraction detector width in the EDAX convention). Since the projection center corresponds to the location on the pattern from which a line drawn to the source of diffraction is normal to the detector, the crystal direction in the detector normal IPF corresponds to the zone axis that appears at the projection center. Fig. 3b shows the pattern for one of the points with a near maximal CI of 0.0934. Here, the [001] zone lies precisely at the center of the pattern, and only one <111> zone is visible at the top edge of the pattern. Fig. 3c shows the pattern with maximal CI=0.0958, which also features the [001] zone in the middle of the pattern. However, this pattern is rotated ~45° compared to Fig. 3b such that the ambiguous <111> zones are just off of the pattern (top-left and top-right corners), which may explain the slightly higher CI.

In Fig. 4, we analyze the effect of rotations about the detector normal in more detail for the highest CI pattern (Fig. 3c). As shown in Fig. 4a, CI varies by a factor of two depending on the rotation angle. As the rotation angle increases, CI sharply drops from 0.0958 before reaching a minimum at ~0.046 near 120° and then slightly increasing to ~0.054 at 180°. The plot appears to be symmetric about 0/360° and 180° with sharp cusps at these rotation values. The small amount of noise visible is likely an artifact of the orientation optimization procedure, suggesting an error in CI of ~0.001. To help understand these results, Fig. 4b shows the corresponding patterns for a few different rotation increments. As the rotation angle increases, the pattern rotates about the projection center (not the center point of the pattern). It is apparent that the 60/300° and 120/240° patterns are mirror images of each other, which explains the symmetry observed in Fig. 4a; this occurs for these specific patterns because the projection center lies on a mirror line of the pattern but is not generally the case. At the minimum CI rotation of 120/240°, the distinguishing [001] zone is near the edge of the pattern whereas the ambiguous [$\bar{1}$11] zone is prominently featured in the bottom half of the pattern. This contrasts with the highest CI rotation of 0/360°, where the [001] zone is at the center of the pattern, and the <111> zones are located just off the pattern.

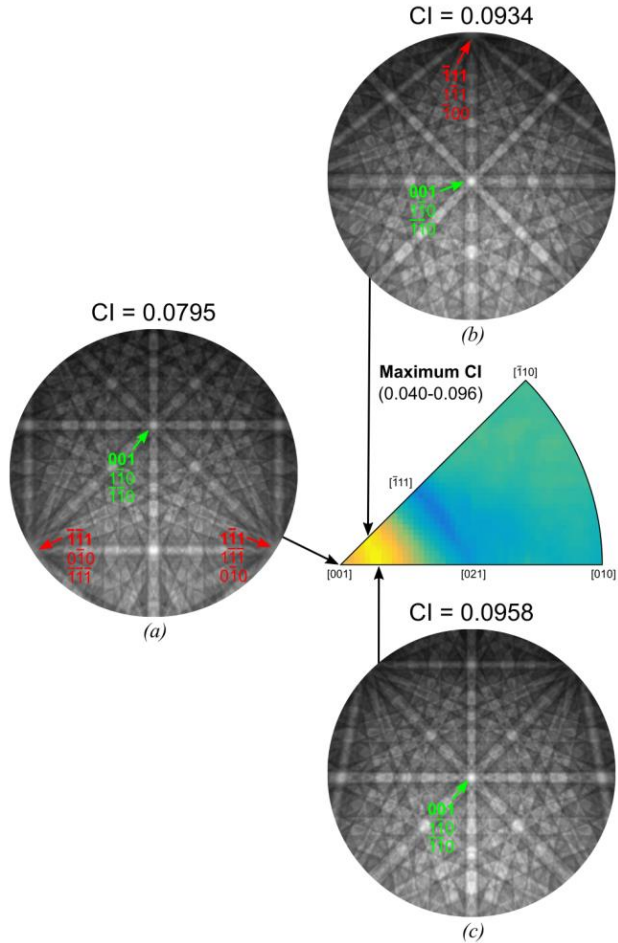

**Fig. 3.** Comparison of EBSD patterns for the (a) [001] orientation with (b-c) two patterns slightly away from [001] with a higher CI. For indexed zone axes, the top label in bold is the correct indexing whereas the lower labels are indexings for the two pseudosymmetry variants. Crystallographically unique zone axes are labeled in green, whereas ambiguous zone axes are labeled in red.

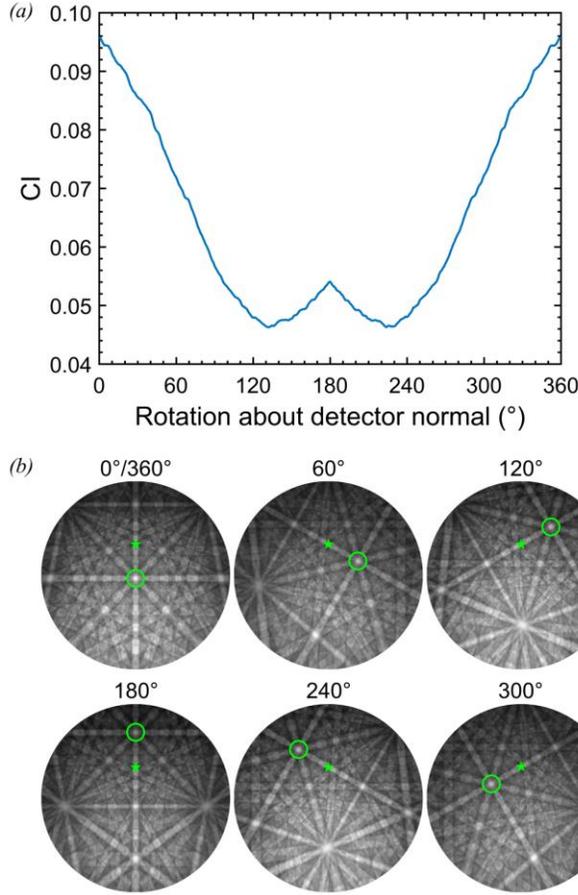

**Fig. 4.** Effect of rotation about the detector normal for the highest CI orientation. (a) CI vs. rotation. (b) Corresponding patterns. The location of the projection center is denoted by a star. The crystallographically unique [001] zone axis is circled.

*2.3. Discussion*

The results from Fig. 3 and Fig. 4 clearly show that the highest CI orientations have the crystallographically unique [001] zone precisely at the center of the pattern. The fact that the center of the pattern is not coincident with the projection center explains why there is a band of high CI values slightly away from [001] on the detector normal IPF. We note that we chose a $Y^*$ value greater than 0 px (above the center) because this is a preferred experimental detector geometry that gives greater diffracted intensity compared to $Y^*=0$ px. When $Y^*\neq0$ px (and $X^*=0$ px), we note that a crystal rotation of $\tan^{-1}(Y^*/Z^*)$ about the horizontal axis would move the [001] zone from the projection center to the middle of the pattern. For the projection center used here ($X^*=0$ px, $Y^*=80$ px, $Z^*=15000$ μm, with a 50 μm pixel size), this gives a rotation of ~15° away from [001], which well describes the observed band of high CI values. When $X^*\neq 0$ px or $Y^*\neq 0$ px (virtually all experimental setups), CI is not only dependent on the detector normal direction but also rotations about the detector normal; therefore, the full orientation dependence of CI cannot be captured by a single IPF (although in many cases a single IPF of the detector normal direction can still be useful to display the data, as shown in Fig. 1). In the case where $X^*=Y^*=0$ px, the highest CI values would be expected to appear precisely at [001] on the detector normal IPF. In addition, the rotation dependence about the detector normal shown in Fig. 4 would be expected to disappear. In such a case, rotations of the crystal about the detector normal would be equivalent to rotating the pattern about its center point (which is coincident with the projection center); no features would move on or off the circular pattern, and a single IPF of the detector normal direction would therefore be able to capture the full orientation dependence of CI.

The precise locations at which the zones appear within the pattern appears to be highly important for intensity-based indexing. As shown in Fig. 3, shifting the [001] zone upwards from the center of the pattern by 80 px (1/6 of the pattern height) leads to a considerable drop in CI from 0.0958 to 0.0795. This arises because there are considerable intensity differences in the region around the [001] zone that can be used to distinguish between pseudosymmetry variants as seen in Fig. 1e; as parts of these regions move off the pattern, the CI correspondingly decreases. It is believed that intensity-based indexing is more sensitive to these small shifts in zone positions and is thus more strongly orientation dependent. Since Hough-based indexing only uses band position information, the CI would not be expected to change unless certain bands moved completely off the screen (or at least far enough such that it was not selected as one of the most prominent bands/reflectors by the indexing algorithm). This contrasts with the case of intensity-based indexing where slight shifts in the zone positions move regions of distinguishing or ambiguous information around that zone on or off the screen, which will affect CI even if the same bands appear in the pattern.

The orientation dependence demonstrated in Fig. 2 is considerably different than that shown by Nowell and Wright for Hough-based indexing (Nowell & Wright, 2005). While they predicted near-zero CI values for all orientations (highlighting the overall difficulty Hough-based indexing methods have with tetragonal $ZrO_2$), they predicted the lowest CI near $[\bar{1}10]$ and the highest CI between $[\bar{1}11]$ and [021]. This may have been partially a result of differences in projection center (they used $Y^*=0$ px) or their choice of reflectors but perhaps also suggests an inherent difference between the indexing approaches: while strong intensity differences around the [001] zone axis can be exploited to clearly distinguish between [001] and [110] zone axes by intensity-based indexing methods, the band positions may not have such a clear distinction that can be detected by Hough-based indexing approaches (at least within the error of the band detection step).

## 3. Effect of detector distance

### 3.1. Understanding trends in CI

To explore the effect of detector distance $Z^*$ on our ability to resolve pseudosymmetry, we simulated patterns from four selected orientations (Fig. 5a) with varying $Z^*$ (while maintaining $X^*=0$ px and $Y^*=80$ px) and computed the CI of the simulated patterns using the same procedure as in Section 2.2. Fig. 5b shows how CI varies with detector distances between 10000 μm (solid angle of 96.7° or 2.58 sr) and 30000 μm (43.0° or 0.55 sr) for these five orientations. In each case, CI decreases as the detector moves further away from the sample, illustrating the clear benefit of capturing a larger solid angle. However, the rate at which CI decreases differs for the four different orientations. In addition, orientations #2 and #4 show a bend in the plot whereas the other orientations exhibit a more constant rate of decline.

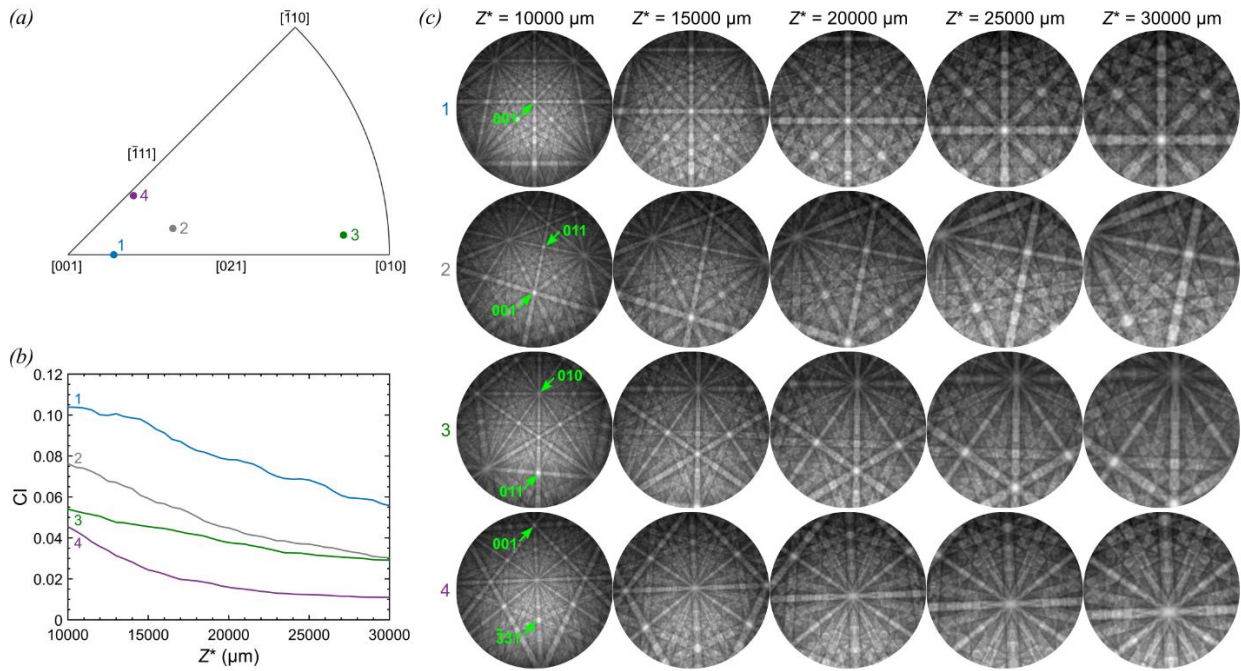

**Fig. 5.** Effect of detector distance $Z^*$ on CI for four selected patterns. (a) IPF of the detector normal direction. (b) Variation of CI with $Z^*$, illustrating that some orientations are more sensitive to detector distance. (c) Patterns showing the magnification effect at higher $Z^*$ values. Main crystallographically unique zone axes are labeled.

In Fig. 5c, we show patterns for these orientations at various detector distances to understand these trends in CI. In each case, as the detector distance increases (and the solid angle decreases), the pattern zooms in about the projection center. For orientation #2, we note that the crystallographically unique [011] and [001] zones are featured in the pattern at $Z^*$=10,000 μm. As $Z^*$ increases, the [001] zone moves downward in the pattern until it disappears just above 20,000 μm; this coincides with the bend in the CI curve after which CI decreases less rapidly. A similar effect was observed for orientation #4, where the [001] zone shifts off the pattern at about 15,000 μm, again corresponding with the bend in the CI curve. Further confirming the importance of the [001] zone, orientation #3, whose patterns never contain the [001] zone, exhibits a constant low rate of CI decline whereas orientation #1, whose patterns always contain the [001] zone, exhibits a constant high rate of CI decline. We also note that, a unique [011] zone moves off the pattern at about 15,000 μm for orientation #3, and a unique [$\bar{3}$31] zone moves off the screen at around 20,000 μm for orientation #4. In both cases, no clear bend in the CI plot can be seen at these $Z^*$ values. These observations demonstrate the importance of the [001] zone specifically and not just any crystallographically unique zone: when the [001] zone is located within the pattern, CI more rapidly declines with increasing $Z^*$ because the magnification effect on the pattern leads to distinguishing regions surrounding the [001] zone disappearing from the pattern. When the [001] zone is not present in the pattern, CI more slowly declines with increasing $Z^*$ as the regions of orientation space being discarded contain less uniquely identifying information.

*3.2. Fitted projection center accuracy*

Paramount to the accuracy of any indexing method is the fitting of an accurate projection center (Pang *et al.*, 2020*a*; Ram *et al.*, 2017). In this section, we characterize the effect of detector distance on our ability to fit accurate projection centers. We used the *pcglobal* package (Pang *et al.*, 2020*a*) with the same parameters and starting points as in (Pang *et al.*, 2020*b*) and fit to 10 noise-free simulated patterns (same parameters as throughout this paper) of random orientations. Fig. 6a shows that the average CI increases at shorter detector distances, and the correct pseudosymmetry variant was selected in each case. During this fitting process, both the projection center and orientation are simultaneously optimized (in contrast to all other results in this paper where the projection center is held fixed while orientation is optimized), which leads to lower CI values here compared to Fig. 5, for reasons discussed in (Pang *et al.*, 2020*a*). Fig. 6b-d shows the absolute error in the fitted projection center components $X^*$, $Y^*$, and $Z^*$, which clearly demonstrates that lower errors in all projection center components are obtained when fitting to a single pattern with smaller $Z^*$. The averaged projection centers, which incorporate error cancelling from positive and negative errors, are given in Table 1.

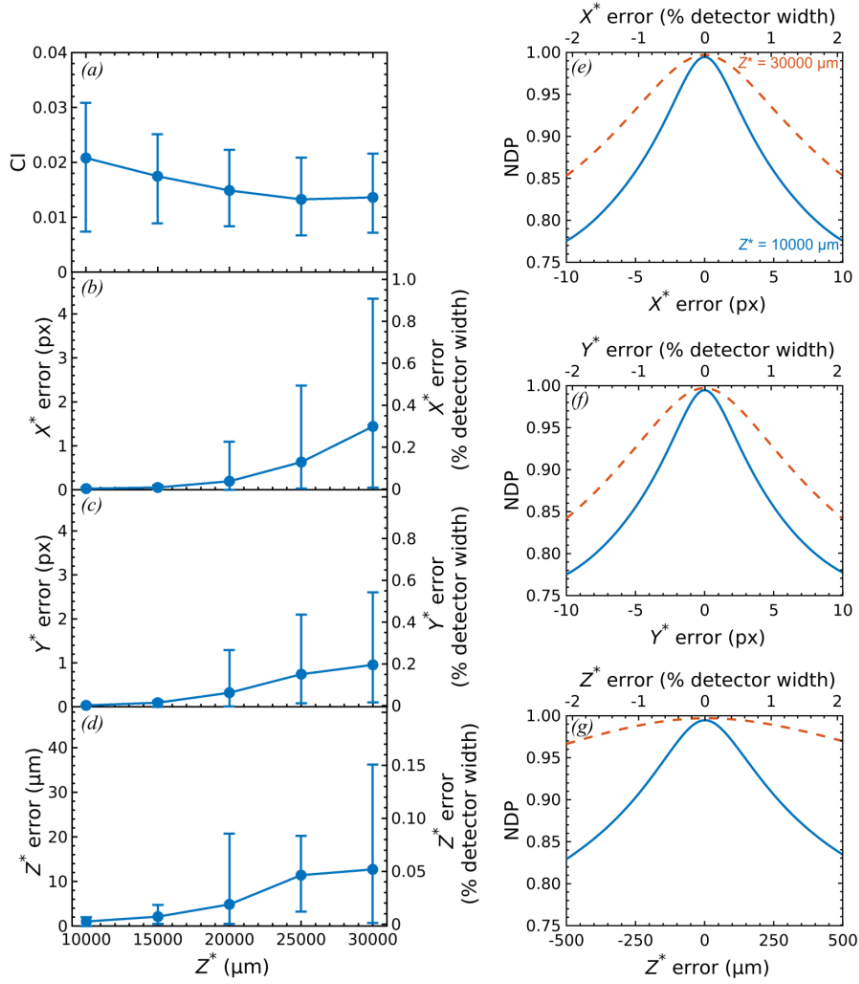

**Fig. 6.** Effect of detector distance $Z^*$ on the fitted projection center accuracy. (a) CI and (b-d) absolute errors in $X^*$, $Y^*$, and $Z^*$ fitted using the *pcglobal* package. Points represent the mean over 10 randomly oriented patterns, and the error bars represent the minimum and maximum errors. (e-g) Linescans showing the variation in NDP with errors in $X^*$, $Y^*$, and $Z^*$ for $Z^*$=10000 and 30000 μm.

**Table 1.** Mean and standard deviation of the pattern center fitted using the *pcglobal* package on 10 randomly oriented patterns at varying detector distances.

| Detector distance | $X^*$ | $Y^*$ | $Z^*$ |
| --- | --- | --- | --- |
| 10000 μm | -0.009 ± 0.034 px | 79.978 ± 0.029 px | 10001.0 ± 1.0 μm |
| 15000 μm | -0.033 ± 0.055 px | 79.953 ± 0.094 px | 15001.7 ± 2.0 μm |
| 20000 μm | 0.068 ± 0.384 px | 80.123 ± 0.504 px | 19999.3 ± 7.8 μm |
| 25000 μm | -0.043 ± 1.006 px | 79.734 ± 1.055 px | 25002.7 ± 13.3 μm |
| 30000 μm | -0.480 ± 2.287 px | 80.312 ± 1.310 px | 30001.1 ± 17.3 μm |

To reveal why more accurate projection centers are obtained at shorter detector distances, in Fig. 6e-g we plot linescans through the optimization landscape in the $X^*$, $Y^*$, and $Z^*$ directions. These linescans demonstrate that the NDP peak is narrower and more isotropic (similar peak widths in the three directions) at smaller $Z^*$, both characteristics that allow optimization algorithms to locate the optimum more accurately.

### 3.3. Results from modified dictionary indexing

Using the projection centers found in Table 1, we indexed patterns for the same 100 orientations from Fig. 1 with various $Z^*$ values (maintaining $X^*=0$ px and $Y^*=80$ px) using the modified dictionary indexing method (same parameters). Compared to the results shown in Fig. 5, this approach takes into account practical sources of error from projection center fitting and interpolation error from the modified dictionary indexing method and gives a more realistic estimate (but still an upper performance bound since we are using simulated patterns) of our ability to resolve pseudosymmetry in tetragonal $ZrO_2$ (c/a=1.0185) at various detector distances. The resulting CI values are shown in an IPF of the detector normal direction in Fig. 7. The same orientation dependence is observed in all cases, but a clear decrease in CI can be observed with increasing $Z^*$ as expected. For $Z^*<30,000$ μm, the correct pseudosymmetry variant was selected for all 100 patterns. However, for $Z^*=30,000$ μm, the incorrect variant was selected for two patterns; at such large detector distances, the small portion of orientation space captured in the pattern is insufficient to confidently distinguish between pseudosymmetry variants for certain orientations after consideration of practical error limitations. Although, we note that the interpolation accuracy of the modified dictionary indexing method can be improved by a finer grid resolution with an associated computational cost, which would likely resolve these errors.

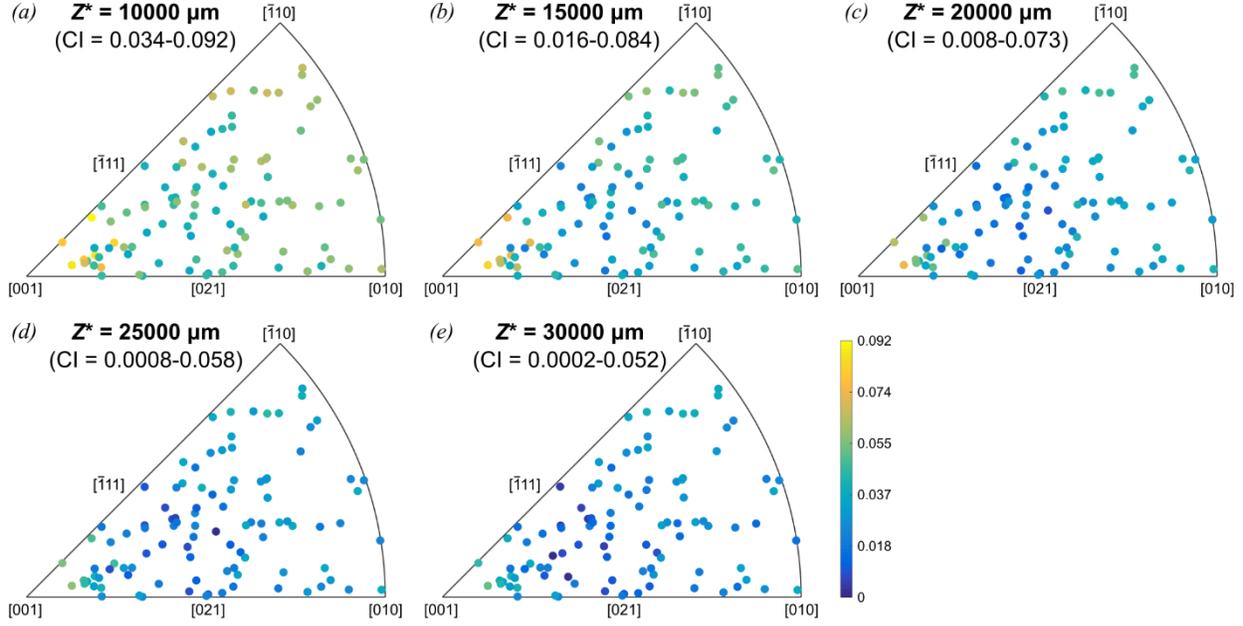

**Fig. 7.** Effect of detector distance $Z^*$ on the CI of 100 random orientations. IPFs are given with respect to the detector normal direction.

Fig. 8a shows the distribution of disorientations between the actual and indexed orientations of the 100 patterns. The data clearly show that more accurate orientations are obtained at shorter detector distances. To understand why, we look at linescans through orientation space to get an idea of how $Z^*$ affects the optimization landscape (Fig. 8b). As can be seen, large $Z^*$ values make the NDP peak narrower in two directions and more anisotropic, both of which make the final interpolation step less accurate (Pang *et al.*, 2020*b*) and thus broaden the disorientation distribution seen in Fig. 8a. Another contributing factor is the larger projection center errors at larger $Z^*$ values, which lead to compensating rotations that bias the indexed orientations and shift the disorientation distribution to the right (Ram *et al.*, 2017; Pang *et al.*, 2020*a*).

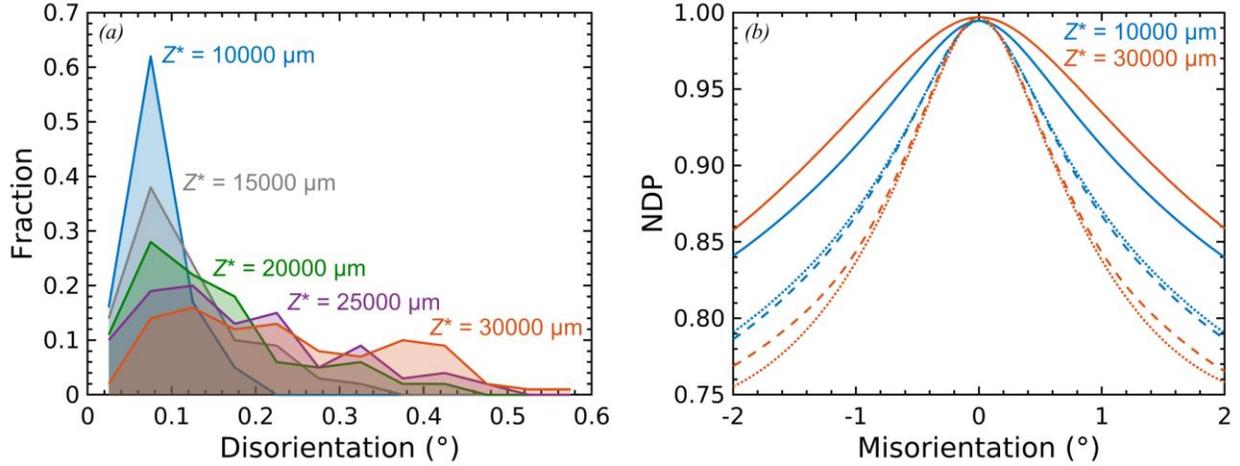

**Fig. 8.** Effect of detector distance $Z^*$ on the orientation accuracy. (a) Distribution of disorientations between the indexed and actual orientations of 100 randomly oriented patterns indexed using the modified dictionary indexing method. (b) Linescans showing the variation in NDP along three orthogonal directions in Rodrigues orientation space (solid, dashed, and dotted lines) for $Z^*$=10000 and 30000 μm.

*3.4. Discussion*

Our modified dictionary indexing results (Fig. 7), as well as our analysis of the theoretical CI in Fig. 5b, clearly show that CI generally increases at shorter detector distances for all crystal orientations. The reason is that at lower $Z^*$ values, a significantly larger portion of orientation space is sampled compared to higher $Z^*$ values; these regions of orientation space that are captured in patterns with lower $Z^*$ values contain intensity information that can uniquely identify between pseudosymmetric orientations (although certain regions, e.g. around <001> zones, contain more distinguishing information than others). To further quantify this, patterns with our detector geometry and $Z^*$=10,000 μm covered a solid angle of 2.58 sr, almost five times more than patterns with $Z^*$=30,000 μm (solid angle of 0.55 sr). Correspondingly, we found a significantly higher CI of 0.034-0.092 for $Z^*$=10,000 μm compared to 0.0002-0.052 for $Z^*$=30,000 μm. This quantitatively validates the assumption made by Lenthe et al. that shorter detector distances aid in resolving pseudosymmetry for intensity-based indexing methods (Lenthe *et al.*, 2019). These results also support the idea that pseudosymmetry is more easily resolved with shorter detector distances for Hough-based indexing because of the larger solid angle sampled (Ryde, 2006; Karthikeyan *et al.*, 2013), although this requires additional verification because of the fundamental differences between Hough-based and intensity-based indexing approaches.

CI does not appear to be more orientation dependent at larger $Z^*$ values, which might be intuitively expected since the patterns cover a smaller solid angle. From the CI ranges computed from the min/max CI values given in Fig. 7, we see a non-monotonic trend: 0.058, 0.068, 0.065, 0.057, and 0.052 in order of increasing $Z^*$. A similar trend is seen in Fig. 5b, with the CI ranges (between orientations #1 and #4) for the same $Z^*$ values being 0.058, 0.071, 0.062, 0.056, and 0.045. It thus appears that detector distances near 15000 μm (0.625 in fraction detector width in the EDAX/TSL convention) lead to the strongest orientation dependence of CI. The weaker orientation dependence going from $Z^*$=15,000 μm to $Z^*$=10,000 μm agrees with our intuition, as patterns now cover such a large portion of orientation space that they are likely to contain highly distinguishing information no matter the crystal orientation. In fact, if $Z^*$ were further decreased, the CI values for the different orientations would be expected to converge. Furthermore, if all of orientation space could be captured in a single pattern, then CI would not be orientation dependent since the orientation dependence fundamentally derives from the inability to capture all of orientation space.

The weaker orientation dependence at $Z^*$ values above 15,000 μm can be rationalized from the results shown in Fig. 5. For the lowest CI orientation #4, the CI begins to plateau at large $Z^*$ values because the pattern is zooming in on an ambiguous <111> zone. There is a certain radius about this zone that contains minimal distinguishing information, e.g. the pattern shown in Fig. 5c for $Z^*$=20,000 μm. Removing information from the edge of this pattern by moving the detector further away to $Z^*$=30,000 μm leads to minimal loss of distinguishing information and thus little change in CI. However, this is not the case for the highest CI orientation #1, where the patterns zoom about a distinguishing [001] zone. Here, CI exhibits a near constant rate of decline as $Z^*$ increases because there is always distinguishing information being lost as regions around the [001] zone are discarded when $Z^*$ decreases (at least in the range studied), as regions quite far from the [001] zone contain a similar density of distinguishing information as regions close to the [001] zone. These differing trends for the limiting orientations, which relate to the range of the distinguishing information about the <001> and <111> zones, lead to the reduced spread in CI at larger detector distances.

It has been suggested, at least for Hough-based indexing, that while shorter detector distances allow for pseudosymmetry to be more easily resolved, it comes with a drawback of reduced orientation accuracy (Karthikeyan *et al.*, 2013). Using an intensity-based modified dictionary indexing method, however, we did not observe any trade-off between these two quantities. In fact, we have found that pseudosymmetry is more easily resolved (Fig. 7) and more accurate orientations are obtained (Fig. 8a) at shorter detector distances. We note that the modified dictionary indexing method used here differs from other intensity-based indexing methods that directly optimize (rather than interpolate) the orientation (Nolze *et al.*, 2018; Singh *et al.*, 2017). Using the SNOBFIT global optimization algorithm to directly optimize the orientations, we still found larger disorientations at longer detector distances (averages of 0.007° and 0.052° for $Z^*$=10000 and 30000 μm, respectively). In this case, the trend was driven by larger projection center errors at larger $Z^*$ values; for a constant projection center error, direct optimization leads to slightly lower disorientations at higher $Z^*$ values as suggested, although this is unattainable in practice because of the need to fit the projection center.

While these results point to the benefits of using the shortest detector distance possible, there are practical considerations that limit how close the detector can be placed to the sample. If the extension of the sample surface impinges upon the detector, then the sample will cast a shadow on the detector and no diffraction pattern will be obtained below this line; this leads to a loss of information, which likely will have negative effects on orientation accuracy and ability to resolve pseudosymmetry. In addition, close detector distances require a small sample size and also run a higher risk of the sample hitting and damaging the detector. Nonetheless, it appears that placing the detector as close as safely possible to the sample without casting a shadow maximizes the likelihood of correctly resolving pseudosymmetry and minimizes the orientation error using intensity-based indexing methods.

## 4. Conclusions

In this work, we studied the effect of crystal orientation and detector distance on the ability to resolve pseudosymmetry in tetragonal $ZrO_2$ (c/a=1.0185) using intensity-based indexing approaches. We found that the orientations having the highest confidence index (CI) in resolving pseudosymmetry have the detector normal nearly parallel to the <001> crystal direction, and the orientations with the lowest CI have the detector normal between the <111> and <012> directions. When the projection center is not located precisely at the center of the pattern, the highest CI orientations are found slightly away from the <001> direction, and the orientation dependence also becomes more complex as rotations about the detector normal must also be considered. Regions around <001> zones are the most useful to distinguish between pseudosymmetry variants, whereas regions around <111> zones are the least useful. As such, the orientations giving the highest CI have an <001> zone located at the center of the pattern, whereas the orientations having the lowest CI have <111> zones prominently featured in the pattern. Small position changes of these zones within the pattern can significantly affect CI, possibly making intensity-based indexing approaches more sensitive to orientation than Hough-based indexing methods.

We also characterized the effect of detector distance in the range 10,000 μm ($Z^*=0.416667$ in the EDAX/TSL convention) to 30,000 μm ($Z^*=1.25$). In general, CI increases at shorter detector distances for all orientations, although the exact nature in which CI changes with detector distance is orientation dependent and can be related to the positioning of the <001> and <111> zones within the pattern. In addition, using a modified dictionary indexing method, we found no trade-off between CI and orientation accuracy as previously suggested for Hough-based indexing, with higher CI and more accurate orientations both being obtained at shorter detector distances. These findings point to the clear benefits of shorter detector distances as an acquisition strategy to obtain more accurate EBSD data with intensity-based indexing approaches.

**Acknowledgements**

The authors would like to thank Dr. Stuart Wright for useful discussions and Dr. Peter Larsen for helpful feedback on the manuscript.

**Funding information**

This material is based upon work sponsored in part by the U.S. Army Research Office through the Institute for Soldier Nanotechnologies, under Cooperative Agreement number W911NF-18-2-0048.

Financial support is also acknowledged from the NSF Graduate Research Fellowship Program under grant number DGE-1745302 (E.L.P.).